\documentclass[11pt]{article}
\usepackage[margin=1in]{geometry}
\usepackage[T1]{fontenc}
\usepackage{lmodern}
\usepackage{amsmath}
\usepackage{graphicx}
\usepackage{booktabs}
\usepackage{tabularx}
\usepackage{caption}
\usepackage{microtype}
\usepackage{xcolor}
\usepackage{xspace}
\usepackage[numbers,sort&compress]{natbib}
\usepackage[hidelinks]{hyperref}
\hypersetup{
  pdftitle={Numbat: Building and Verifying a Self-Contained
    Machine-Learning Stack},
  pdfauthor={Thang Tran, Lan Dang},
}

\graphicspath{{figures/}}
\newcommand{\FinalMap}{0.4956\xspace}      % mAP50-95, FINAL epoch-500 EMA (released)
\newcommand{\FinalMapFifty}{0.6610\xspace} % mAP50,    FINAL epoch-500 EMA (released)

\newcommand{\yolonbm}{YOLO\mbox{-}NB\mbox{-}M\xspace}

\title{\textbf{Numbat: Building and Verifying a Self-Contained\\
Machine-Learning Stack}}
\author{%
  Thang Tran\thanks{Corresponding author: \texttt{thang.tran@cloudkites.com}}\\
  \small CloudKites AI Lab\\
  \small New South Wales, Australia\\
  \small\texttt{thang.tran@cloudkites.com}
  \and
  Lan Dang\\
  \small Monash Business School, Monash University\\
  \small Victoria, Australia\\
  \small\texttt{LanHong.Dang@monash.edu}
}
\date{September 2026}

\begin{document}
\maketitle

\begin{abstract}
Machine-learning systems are built almost exclusively on a few large
Python-orchestrated frameworks, and they inherit those stacks' engineering
costs: environments of hundreds of version-coupled packages, separate export
toolchains for deployment, and the split between the language research is
written in and the language products ship in. We report on the construction
and verification of \emph{numbat}, a machine-learning stack written in one
general-purpose language (Zig) with no third-party runtime dependencies. The
stack spans tensor computation, automatic differentiation, neural-network
modules, mixed precision, multi-GPU training, data loading and monitoring; an
SDK exposes it behind a stable, additively versioned C ABI of over 1{,}400
entry points, with bindings for six languages; and its clinical domain planes
encode regulatory requirements as executable acceptance gates rather than
documentation. Verifying such a stack is the harder half of building it: a
defective training run rarely fails, it converges quietly to a slightly worse
model. We treat a widely used reference implementation as an executable
specification and verify against it at five levels, from operator gradient
checks to an automated trajectory gate against a same-machine reference run
--- the arrangement our companion study formalizes as a trajectory-level
differential oracle. The protocol surfaced ten silent recipe divergences,
which we catalog with mechanisms and symptoms. As the acceptance test, we
train a 25.9M-parameter detector of the YOLOv8m class from random
initialization on COCO~2017 for the full 500-epoch schedule: the exported
weights score \FinalMap{} mAP$_{50\text{--}95}$ under the official protocol,
scored by the reference stack's own validator (published endpoint 0.502),
with single-GPU step time at parity on identical hardware. Weights, per-epoch
metrics and the full run manifest are released.
\end{abstract}

\section{Introduction}\label{sec:intro}

Contemporary deep-learning practice is concentrated on a small number of large
Python-orchestrated frameworks~\citep{paszke2019pytorch,abadi2016tensorflow,
bradbury2018jax}. These systems are mature and productive. They are also
expensive in ways their own authors acknowledge~\citep{paszke2019pytorch,
ansel2024pytorch2}. Per-operator dispatch through the interpreter adds
host-side overhead, which PyTorch~2 mitigates with an additional
bytecode-capture and graph-compilation layer~\citep{ansel2024pytorch2} at the
price of further system complexity. A GPU-enabled installation spans gigabytes
across hundreds of packages, and the version coupling among framework,
accelerator libraries, driver and Python dependencies makes environments
costly to reproduce and audit. Deployment to edge, embedded, mobile or
air-gapped targets goes through separate export toolchains rather than through
the stack that trained the model, yielding the familiar two-language pattern
--- research in Python, production re-implemented in a systems language ---
which duplicates engineering and invites divergence. Organizations that must
audit their supply chain, or bound memory behavior on constrained devices,
inherit all of it. The demand for self-contained runtimes is directly
evidenced by the adoption of inference engines such as
\texttt{llama.cpp}~\citep{gerganov2023llamacpp} and ONNX
Runtime~\citep{onnxruntime2018}; such systems deliberately stop short of
\emph{training}. Independent training stacks outside the mainstream ecosystem
exist~\citep{burn2024,tinygrad2023,hannun2023mlx}, but public evidence that
one can reproduce the \emph{training outcome} of a mature, competitive recipe
--- rather than merely execute gradient descent --- remains scarce.

numbat removes these costs at their root: one dependency-free stack in a
single general-purpose language, covering training through deployment, with
applications shipping as single static binaries. Building it, however, is only
half the work, and the smaller half. A modern training recipe is specified
\emph{de facto} by its implementation rather than by any paper: initializer
distributions, mixed-precision autocast placement, optimizer parameter
grouping, exponential-moving-average (EMA) semantics, gradient-clipping
policy, loss-scaler dynamics, augmentation details and random-number-generator
(RNG) structure are behavioral properties of code, individually undocumented
and individually capable of degrading the final metric without any overt
failure. This is a recognized reproducibility problem even \emph{within} one
framework~\citep{pineau2021reproducibility,henderson2018deeprl,
bouthillier2019unreproducible}; across independently implemented frameworks it
is the central difficulty, and it is a software-engineering difficulty ---
a verification problem with no oracle --- before it is a numerical one. In a
companion study we formalized the two-stack arrangement as a trajectory-level
differential oracle and applied it to language-model fine-tuning, where the
comparison exposed seventeen faults that single-implementation development had
missed~\citep{tran2026crossstack}. This paper is about the stack itself: how
it is put together, how its interfaces are kept stable, how domain
requirements are made executable, and what it took to reproduce a competitive
from-scratch training outcome end to end.

The paper makes five contributions.

\begin{enumerate}
  \item \textbf{Design.} We describe numbat, a self-contained ML framework
  written in Zig covering tensors, autograd, modules, mixed precision,
  distributed data-parallel training, data pipelines and monitoring, with CPU
  and CUDA backends in production use and vendor-neutral GPU paths including a
  single-source kernel plane compiled by the framework's own compiler
  (\S\ref{sec:framework}). One design principle carries most of the weight:
  \emph{reference-default semantics} --- numbat adopts the dominant
  ecosystem's defaults (initializers, autocast policy, optimizer and
  loss-scaler defaults, RNG streams) as its own, so recipes transfer across
  stacks unchanged.
  \item \textbf{Interface discipline.} The numbat SDK packages the full
  training and inference capability behind one C ABI that is versioned and
  strictly additive, with thin bindings for six languages whose numerical
  agreement is enforced by shared tests and, at trajectory level, evidenced in
  the companion study (\S\ref{sec:sdk}).
  \item \textbf{Requirements as executable gates.} For the clinical domains
  the stack targets, design requirements are encoded as acceptance gates ---
  executable checks with verdicts and exit codes --- backed by typed
  diagnostic registries whose coverage is measured rather than asserted
  (\S\ref{sec:planes}). We describe the method and what it caught.
  \item \textbf{Reproduction protocol and divergence catalog.} We define a
  five-level cross-framework training-reproduction protocol --- operator,
  module, step, trajectory, outcome --- with an automated trajectory gate
  against a same-machine reference trajectory (\S\ref{sec:method}), and we
  catalog the ten silent recipe divergences it surfaced, with mechanisms and
  symptoms (\S\ref{sec:divergences}). The catalog is of independent value to
  anyone porting a training recipe between frameworks.
  \item \textbf{End-to-end evidence.} We train \yolonbm, a YOLOv8m-class
  detector, from scratch on COCO~2017 for 500 epochs on three power-capped
  consumer GPUs, tracking the reference trajectory throughout, and report
  accuracy under the official COCO protocol as scored by the \emph{reference
  stack's own validator} on exported weights, plus same-machine throughput,
  memory and energy accounting, including negative results
  (\S\ref{sec:results}). Weights, metrics and the run manifest are released
  (\S\ref{sec:repro}).
\end{enumerate}

\section{Related work}\label{sec:related}

\paragraph{Frameworks and runtimes.} PyTorch~\citep{paszke2019pytorch},
TensorFlow~\citep{abadi2016tensorflow}, and JAX~\citep{bradbury2018jax}
dominate training workloads and anchor a large ecosystem of Python packages.
Outside this ecosystem, MLX~\citep{hannun2023mlx} targets Apple silicon,
Burn~\citep{burn2024} provides a Rust-native training stack, and
tinygrad~\citep{tinygrad2023} explores a minimal operator set; llama.cpp/ggml
\citep{gerganov2023llamacpp} and ONNX Runtime~\citep{onnxruntime2018}
demonstrate the deployment value of self-contained \emph{inference}. numbat
differs in combining (i) a single-language, dependency-free implementation of
the \emph{full training stack}, (ii) explicit behavioral compatibility with
PyTorch defaults, and (iii) a packaged SDK exposing that stack from six
languages behind one stable ABI (\S\ref{sec:sdk}).

\paragraph{The YOLO family.} Single-stage real-time detection descends from
YOLO~\citep{redmon2016yolo,redmon2018yolov3} through successive architecture
generations~\citep{bochkovskiy2020yolov4,li2022yolov6,wang2023yolov7,
jocher2023yolov8}. The eighth-generation ``v8'' class combines a cross-stage-partial (CSP)
backbone~\citep{wang2020cspnet}, a path-aggregation feature-pyramid (PAN-FPN)
neck~\citep{lin2017fpn,liu2018panet},
an anchor-free decoupled head with distribution focal loss (DFL) box
regression~\citep{li2020gfl}, task-aligned label assignment~\citep{feng2021tood},
and complete-intersection-over-union (CIoU) box
loss~\citep{zheng2020diou}. We train a detector of this
architecture class (25.9M parameters at medium scale) as our case study.

\paragraph{Training detectors from scratch.} DSOD~\citep{shen2017dsod} and
subsequent analysis~\citep{he2019rethinking} established that detectors can be
trained from random initialization to pretrained-equivalent accuracy given
sufficient schedule length; modern YOLO recipes are routinely trained from
scratch. Our contribution is orthogonal: not \emph{whether} from-scratch
training works, but whether an \emph{independent framework} can reproduce a
reference recipe's outcome exactly, and what silently breaks along the way.

\paragraph{Reproducibility and cross-stack validation.} Sensitivity of
outcomes to seemingly minor implementation details is well
documented~\citep{henderson2018deeprl,bouthillier2019unreproducible,
pineau2021reproducibility}. We extend this line of work to the cross-framework
setting, where the reference implementation must be treated as the
specification, and provide an operational protocol for verifying
convergence-level --- not merely operator-level --- parity. The companion
study~\citep{tran2026crossstack} formalizes the underlying idea: two
independently implemented stacks running one specification act as differential
oracles for each other at the level of the learning trajectory, addressing the
oracle problem that makes training pipelines hard to test at all. There the
workload was language-model fine-tuning and the second stack was numbat driven
through its own SDK; here the same discipline is applied to a from-scratch
detector run, and to the construction of the stack itself.

\section{The numbat framework}\label{sec:framework}

\begin{figure}[t]
  \centering
  \includegraphics[width=\textwidth]{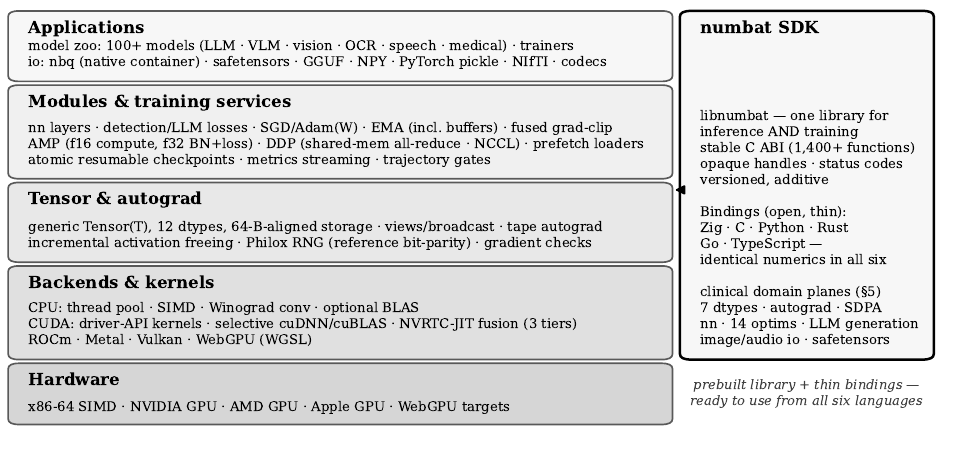}
  \caption{The numbat stack and the numbat SDK. The framework is one
  vertically integrated codebase in a single general-purpose language (Zig);
  the SDK
  (\S\ref{sec:sdk}) packages its full capability as prebuilt libraries behind
  a stable, versioned C application binary interface (ABI) with thin open
  bindings, so applications in six
  languages build on the framework directly.}
  \label{fig:stack}
\end{figure}

numbat is written in Zig.\footnote{\url{https://ziglang.org}} The language
choice is doing real work, not signalling: memory is managed explicitly
through caller-supplied allocators, there is no hidden runtime and nothing
that can pause a training step to collect garbage, and the compile-time
metaprogramming is strong enough that the entire tensor layer is built with it
rather than with a code generator. Figure~\ref{fig:stack} lays the stack out.
Four decisions shaped it; each is stated below along with what it buys.

\emph{(i) Single-language vertical integration.} Tensor storage, kernels,
autograd, modules, optimizers, distributed training, data decoding
(including image codecs), tokenization, and monitoring are one codebase with
no foreign-function seams except vendor GPU libraries. Applications ship as
single static binaries.

\emph{(ii) Explicit, deterministic memory.} All allocations flow through
caller-supplied allocators; tensor storage is 64-byte aligned; lifetimes are
explicit. During backpropagation, saved activations can be freed incrementally
as the tape drains; on the case study of \S\ref{sec:method} this reduces peak
training memory per GPU from 18.5 to 13.5\,GiB ($-26.8\%$) with bitwise
identical gradients.

\emph{(iii) Compile-time specialization.} The tensor type is generic over its
element: \texttt{Tensor(T)}, where \texttt{T} ranges over twelve types, from
booleans and the integer widths up to the four floating-point precisions
(F16, BF16, F32, F64). Both the tensor and the backend dispatch are
monomorphized when the program compiles. The consequence matters more than
the mechanism: at run time nothing stands between model code and the kernels
--- no dispatch layer to cross, no interpreter, no graph VM to feed.

\emph{(iv) Reference-default semantics.} Where the dominant ecosystem has a
default --- initializer distributions, AMP autocast placement, optimizer
hyperparameter defaults, loss-scaler constants, RNG algorithms --- numbat
adopts it as \emph{its} default. The counter-based Philox
generator~\citep{salmon2011philox} is bit-compatible with the reference's GPU
uniform sampling (verified by known-answer tests), and the initializer suite
reproduces \texttt{torch.nn.init} distributions exactly. Section
\ref{sec:divergences} shows why this principle is load-bearing: most
convergence divergences we found were violations of it.

\subsection{Backends and kernel generation}

On the CPU, kernels are vectorized --- single instruction, multiple data
(SIMD) --- and scheduled by a work-stealing thread pool; convolutions go
through the Winograd transform~\citep{lavin2016winograd} at the sizes where
it pays, and an external BLAS (Basic Linear Algebra Subprograms) library can
be linked in for the matrix products, though none is required. The CUDA
backend talks to the driver API directly. Vendor libraries are used where
profiling says they win --- cuBLAS throughout, cuDNN for particular
convolutions --- and fused kernels are generated at three tiers: a curated set of hand-written fused kernels; a runtime
code-generator that compiles arbitrary elementwise chains via NVIDIA's
runtime compiler (NVRTC), cached by
operation signature; and a fusion-graph compiler in the spirit of
Triton~\citep{tillet2019triton} that lowers a typed operator directed acyclic
graph (DAG) into two
kernel templates (pure elementwise with broadcast indexing; reduction with
fused pre/post epilogue, covering the normalization family). On a production
connectionist-temporal-classification (CTC) speech-recognition encoder, graph
fusion reduced per-batch latency from
258\,ms to 23\,ms. The ROCm, Vulkan, Metal, and WebGPU backends follow the same
layering. Two of them are vendor-neutral, and they bracket an engineering
progression. The Vulkan backend --- GLSL compiled ahead of time to SPIR-V ---
is correctness-complete across the operator set with automatic differentiation
on both forward and backward, and reaches the device's tensor cores through
half-precision cooperative-matrix multiplication. The newer single-source
kernel plane goes further: device kernels are written in Zig and compiled to
SPIR-V by the same compiler that builds the framework, then dispatched over
the Vulkan runtime that the display driver already provides. The compiled
kernel pack is embedded in the library, so the GPU path requires no vendor
toolkit, no SDK and no shader compiler on the target machine, at build time or
at run time --- the machine needs a display driver and nothing else. The
WebGPU backend is single-precision and inference-oriented, and the Metal
backend is an early stub.

\subsection{Training subsystem}

\emph{Optimization.} Stochastic gradient descent (SGD) and Adam(W) carry
reference-default semantics throughout, down to details that turn out to be
load-bearing: parameter groups (the case study needs the reference's
three-group structure, \S\ref{sec:divergences}), fused multi-tensor update
paths, and gradient accumulation that preserves nominal-batch semantics.
Gradient clipping is worth its own sentence. The global-norm form is a single
fused kernel --- one squared-norm accumulation across all gradients, one
device-to-host readback --- where a naive port performed 243 per-tensor
synchronizations; on the case study that is the difference between 46--71\,ms
and 7\,ms per step.

\emph{Mixed precision.} Automatic mixed precision (AMP) follows reference
autocast
semantics~\citep{micikevicius2018mixed}: F16 compute for convolutions and
matrix products, F32 master weights, and --- critically
(\S\ref{sec:divergences}) --- F32 batch normalization (BN)~\citep{ioffe2015batch}
and loss, with dynamic loss scaling.

\emph{Distributed data parallelism (DDP).} One process per GPU, following
\citet{li2020pytorchddp}; gradients are reduced via a host-staged, sharded
all-reduce over POSIX shared memory with pinned staging buffers and futex-based
barriers, producing rank-order-deterministic, bitwise-reproducible sums. An
overlap engine that reduces buckets concurrently with backpropagation, and a
transport over the NVIDIA Collective Communications Library (NCCL), are
implemented and gated off by default on small-core hosts
(\S\ref{sec:results:throughput} reports the measured reason honestly).

\emph{Data pipeline.} Background prefetch loaders perform decode and
augmentation in worker threads. All image decoding is in-tree and
reference-parity tested; a notable byproduct of parity testing was the
discovery of a quantization-table ordering bug in our JPEG decoder that
silently perturbed every decoded pixel (mean absolute error 9--20 intensity
levels vs.\ libjpeg) --- exactly the class of silent divergence the protocol of
\S\ref{sec:method} exists to catch. After correction and vectorization the
decoder sustains $1.88\times$ its previous throughput and the loader fully
overlaps GPU compute on the case study.

\emph{Monitoring and gating.} Every run streams its metrics, as CSV and
JSONL, to \emph{nbmonitor} --- numbat's own experiment tracker, self-hosted
behind its own accounts and API keys. Functionally it does what the hosted
trackers do: live charts, run search and comparison, per-GPU telemetry down to
power and temperature, logs, media, sweeps. The difference is where it runs.
Training telemetry never leaves the organization, which for clinical work is a
requirement rather than a preference. The more important consumer of the same
stream is not a person at all: a \emph{trajectory gate} sidecar polls each
epoch and can pause or terminate a run that leaves its reference band
(\S\ref{sec:method:gate}), because a dashboard someone glances at twice a day
cannot stop a run that went wrong at 3\,a.m. Checkpoints are atomic and carry
optimizer, EMA, epoch and RNG state, so a run resumed mid-schedule continues
as if it had never stopped.

\subsection{Interoperability and model zoo}

A framework that cannot read the ecosystem's files is an island, so numbat
reads and writes the formats models actually arrive in: safetensors, GGUF with
its k-quantizations, NPY, PyTorch pickle checkpoints, and the medical and
audio formats its applications need (NIfTI-1 among them). Its tokenizer suite
is byte-identical to the reference tokenizer library on validation corpora ---
byte-identical rather than merely compatible, because \S\ref{sec:divergences}
is a catalog of what "merely compatible" costs.

For deployment, numbat defines \emph{.nbq}, its native self-contained model
container --- a novel format developed at CloudKites AI Lab, for which a
patent application is in preparation. A single \texttt{.nbq} file carries
everything a production application loads: the (optionally quantized) weight
buffers already in their device-ready layout, the tokenizer, the model
configuration, and an authenticated, self-describing manifest recording each
tensor's dtype, layout, and role together with the target backend and minimum
device capabilities --- so a runtime can decide compatibility, with a precise
reason on mismatch, before reading any weight data. The payload is
page-aligned and laid out contiguously in load order, so a model starts with
zero repacking or conversion: memory-mapped on CPU (zero-copy) or transferred
to the accelerator in a single host-to-device copy, with per-tensor device
pointers recovered by offset arithmetic.

The model zoo counts more than a hundred complete implementations
(102 at the time of writing) spanning large language models, vision--language
models, vision (classification, detection, segmentation), optical character
recognition (OCR), speech recognition and synthesis, speaker and audio
analysis, video, and a dedicated medical and clinical group (nnU-Net,
MedSAM2-class segmenters, electrocardiogram models, endoscopy detectors);
LLaMA-3, Qwen-3, Whisper, SAM-class segmenters, DINOv3 ViT and the YOLO
family are examples, not an inventory. Each implementation is the architecture
rewritten in Zig, loading the originally published weights and checked against
the original's output. This breadth is exercised in production: numbat is the sole
inference engine of \emph{emu}, a freely distributed, local-first multimodal
desktop application for Windows and
Linux\footnote{\url{https://huggingface.co/cloudkites/emu}} that runs
language, vision--language, OCR, text-to-speech, and speech-to-text models
entirely on the user's machine (CPU or NVIDIA GPU) --- the single-library
deployment model of \S\ref{sec:sdk} operating in the field. Weight-level
interoperability is what the methodology stands on: numbat-trained
weights are exported to safetensors and \emph{evaluated by third-party stacks}
(\S\ref{sec:method:outcome}).

\section{The numbat SDK}\label{sec:sdk}

Adopting a new framework should not require adopting a new toolchain, nor
should the framework's internals become part of every downstream build. The
numbat SDK packages the framework as one ready-to-use library for building
end-to-end machine-learning applications: training, evaluation and inference
live in the same stack, so the code that fits a model in a research prototype
is the code that serves it in production. The library is self-contained and
cross-platform (Linux, macOS and Windows; x86-64 and ARM64; shared and static
artifacts, plus a WebAssembly build for the browser), and the whole platform
matrix is cross-compiled from a single development machine --- a property of
the Zig toolchain rather than of any release infrastructure. This answers the
deployment-side costs catalogued in \S\ref{sec:intro} directly: no interpreter
in the serving path, no multi-gigabyte environment to reproduce, no separate
export toolchain between the stack that trains a model and the stack that
ships it, and no two-language rewrite between prototype and production.

The SDK is structured in three layers (right column of Fig.~\ref{fig:stack}).

\begin{enumerate}
  \item \textbf{Core}: the framework of \S\ref{sec:framework}, built as
  prebuilt shared and static libraries per platform, with a single C header.
  \item \textbf{C ABI}: one library, \texttt{libnumbat}, exposing inference
  \emph{and} training through what is now more than 1{,}400 \texttt{nb\_*}
  entry points --- the tensor and module core, and the domain planes of
  \S\ref{sec:planes}, which account for most of the surface. Only opaque
  handles, runtime-tagged dtypes and status codes cross the boundary: no
  framework headers, no memory-layout assumptions. That is what keeps the
  interface stable. The ABI is versioned and strictly additive --- at revision
  3 at the time of writing, with the two most recent releases adding 282 entry
  points between them without breaking a single existing caller --- so
  applications built today keep working as the framework evolves underneath
  them. Errors are returned as status codes with structured, thread-local
  detail; all handles are caller-owned with explicit destructors.
  \item \textbf{Bindings} (deliberately thin): Zig, C, Python, Rust, Go and
  TypeScript, all over the same ABI. Each presents a PyTorch-shaped API
  (\texttt{numbat} $\approx$ \texttt{torch}, \texttt{numbat.nn},
  \texttt{numbat.optim}) so that ecosystem experience transfers directly. The
  Python binding depends on the Python standard library and nothing else ---
  deliberately not NumPy --- and the other bindings likewise pull no
  third-party packages; the zero-dependency property of the framework extends
  to everything a consumer links.
\end{enumerate}

What crosses the boundary is the working vocabulary of the ecosystem, not a
reduced export subset. Tensors come in seven runtime dtypes with casting and
full autograd control (\texttt{detach}, \texttt{retain\_grad}, no-grad mode);
scaled dot-product attention is differentiable through the boundary, with
causal and additive masks and grouped-/multi-query heads, on CPU and CUDA. The
\texttt{nn} modules run from convolutions through pooling, normalization,
embeddings and activations --- enough to assemble ResNet- and
transformer-class architectures from any binding --- and training is served by
the torch-parity optimizer and scheduler families, gradient clipping, EMA
weight averaging, seeding, \texttt{torch.nn.init}-parity initializers, and
losses from cross-entropy through the margin and ranking families. Data
loading, module save/load, and I/O for safetensors, NumPy, NIfTI, images and
audio round it out. A language-model path runs end to end through the same
boundary: a GGUF-packaged model loads, tokenizes and generates text, with a
complete sampling toolkit, from any of the six languages. And the models in
the SDK's zoo are not opaque handles. Each is composed in the binding's own
language from those same \texttt{nn} primitives, so a developer reads --- and
can change --- the architecture in the language they already use.

Cross-binding numerical agreement is enforced at two levels. A shared
known-answer test fixes a training task whose loss trajectory must be
identical from all six languages, and a parity harness sweeps the operator
surface. At trajectory level, the companion study provides the strongest
evidence: four implementations of one fine-tuning specification --- PyTorch,
numbat driven natively, and the SDK driven from Python and from Zig --- ended
a full training epoch within 0.15\% of one another in held-out cross-entropy,
with 42 paired evaluations differing by 0.134\% on
average~\citep{tran2026crossstack}. Notably, the same study found that four of
the seventeen faults it exposed were reachable only from a language whose
memory model differed from the others'; the six-language surface is a testing
asset, not only a convenience.

The SDK is assembled into versioned releases --- the prebuilt platform
libraries, the header, the binding sources, documentation and worked examples
--- and every release is verified by installing the built artifact and
interrogating it, a step \S\ref{sec:planes} motivates. GPU support is an
opt-in build of the same library, with vendor runtimes loaded dynamically.

\paragraph{Beyond application software.} The same properties carry the SDK
into robotics and long-running agentic systems: real-time vision, speech and
language models behind one C ABI that links directly into C/C++/Rust control
stacks; a single static library with no interpreter or garbage collector,
whose allocator-controlled memory keeps latency predictable on embedded
compute; and cross-compilation to heterogeneous on-device hardware from one
codebase. Because the one library also trains, on-device adaptation ---
fine-tuning a perception model on locally collected data without a round trip
to a training cluster --- is supported by the same binary. \emph{emu}, a
freely distributed local-first desktop application built solely on numbat
(\S\ref{sec:framework}), exercises the deployment model in the field.

\section{Domain planes: requirements as executable gates}\label{sec:planes}

numbat's application focus is medical and clinical AI, and in that setting
most of the engineering is not the model. It is the domain machinery around
it: reading the hospital's imaging and messaging formats, coding findings to
licensed clinical terminologies, and assembling the evidence a regulator will
ask for. The stack carries this machinery in \emph{planes}: standalone
packages, each depending on the framework but buildable and testable on its
own, covering medical imaging and geometry (including the DICOM wire
protocols), genomics and multi-omics (format-parity-tested against the
reference C implementations), clinical language and terminology, evidence and
evaluation, duplex clinical voice, and a machine-readable surface for AI
agents that build against the stack. Together the planes account for most of
the SDK's entry points.

What makes the planes relevant to a software-engineering audience is not the
domain coverage but the method by which each was built, which we believe
generalizes.

\emph{Requirements are executable.} Each plane's design document enumerates
numbered requirements, and each requirement is discharged by an
\emph{acceptance gate}: an executable check with a verdict and an exit code,
run from the build system like a test. A development phase is complete when
its gate passes, and not before. Gates print the clauses they deliberately do
\emph{not} discharge --- requirements awaiting clinical data, elapsed time or
absent hardware --- so a passing gate cannot be misread as covering them.
Across the six planes the suites currently comprise 98 gates and 1{,}944
individual checks.

\emph{Failures are typed.} Each plane ships a diagnostic registry: stable
numbered codes, each carrying a severity class and a typed repair plan, so a
caller branches on \emph{which} requirement was violated rather than parsing a
sentence that will eventually be reworded. The registry's coverage --- does
every registered code have a real check behind it? --- is measured from a
ledger written at the raise sites, not from a hand-maintained list. This
distinction earned its keep: during development the measured-coverage gate
failed three times, each on a code that was registered, documented and
reachable from no check at all. A hand-kept list would have reported full
coverage at each of those moments.

\emph{Interfaces refuse.} Where a domain rule exists, the API enforces it
rather than documenting it. The terminology plane refuses to build an index
over a licensed vocabulary unless the caller supplies a license reference, and
refuses \emph{before reading a single concept row}. The evidence plane has no
code path that yields a bare metric value: a measurement carries the digest of
the run manifest that produced it or is explicitly marked untraceable, and
quoting an untraceable measurement is refused --- across the C ABI as well as
natively, so a consumer in another language cannot route around it. A claim
gate can block a release outright when the evidence on file does not match the
kind of claim being made, naming in the refusal the study design that would
support it.

\emph{The shipped artifact is part of the test surface.} One incident shaped
the release procedure. The evidence plane's ABI reported that it ran fifteen
acceptance gates; the suite runs sixteen. The ABI kept its own hand-maintained
copy of the gate table, the copy had lost an entry, and the pure-C conformance
test had been written against the same copy --- so it asserted fifteen, and
passed. Every check in the repository was green; the defect was found by
installing the released archive, loading the library and asking it how many
gates it had. Both tables now derive from one source under a compile-time
length check, and every release ends by interrogating the built artifact.
The episode independently corroborates the companion study's finding that the
faults that matter often live outside the code paths that testing effort
concentrates on~\citep{tran2026crossstack}.

\section{Case study: \yolonbm{} from scratch on COCO}\label{sec:method}

\subsection{Task and model}

We train \yolonbm\footnote{Named by its creators: \emph{YOLO} for the
real-time detector family the architecture belongs to, \emph{NB} for the
numbat framework, \emph{M} for medium scale. Third-party product names appear
in this paper nominatively, to identify the architecture class and the
reference implementation used for benchmarking; the model, its weights, and
its name are independent work.}, an 80-class one-stage anchor-free detector of
the YOLOv8m architecture class: CSP backbone, PAN-FPN neck, decoupled head
with DFL box regression ($\mathrm{reg\_max}=16$), SiLU
activations~\citep{elfwing2018silu}, detection strides $\{8,16,32\}$
(8{,}400 candidate locations at $640^2$), task-aligned
assignment~\citep{feng2021tood}, and a binary-cross-entropy (BCE)
classification + CIoU
\citep{zheng2020diou} + DFL~\citep{li2020gfl} composite loss; 25.9M parameters.
Training data is COCO~2017 \texttt{train2017} (118{,}287 images);
evaluation is \texttt{val2017} (5{,}000 images) under the COCO
mean-average-precision (mAP) protocol~\citep{lin2014coco}. All
weights start from random initialization with seed~0; no pretrained weights of
any provenance are used at any point.

\subsection{Recipe}

Table~\ref{tab:recipe} lists the complete recipe. It reproduces the reference
implementation's from-scratch schedule, including behaviors that are
\emph{coded rather than configured} in the reference trainer --- loss scaling
by world size, the three-way optimizer parameter grouping, per-step
global-norm clipping at 10.0, unconditional per-step EMA with ramped decay,
batch-scaled weight decay ($\lambda \cdot b\,w/\mathrm{nbs}$ with nominal
batch $\mathrm{nbs}=64$), warmup interpolation per iteration, and mosaic
shutdown for the final ten epochs. Section~\ref{sec:divergences} documents
what happened when any of these was missed.

\begin{table}[t]
\caption{Complete training recipe (identical to the released run manifest).}
\label{tab:recipe}
\centering\small
\begin{tabularx}{\textwidth}{@{}l X@{}}
\toprule
\textbf{Optimization} & SGD, Nesterov momentum 0.937; $\mathrm{lr}_0=0.01$
with linear decay to $0.01\cdot \mathrm{lr}_0$; 3-epoch per-iteration warmup
(momentum $0.8\!\to\!0.937$; bias-group LR from 0.1); weight decay
$5\!\times\!10^{-4}$ scaled by effective batch ($48/64 \Rightarrow 3.75\times
10^{-4}$); three parameter groups (decayed weights; undecayed normalization
gains; undecayed biases with elevated warmup LR); global-norm clip 10.0 every
step; 500 epochs, 2{,}464 steps/epoch, $1.232\times10^{6}$ iterations. \\
\midrule
\textbf{Precision} & AMP: F16 compute for conv/matmul; F32 batch
normalization, loss, and master weights; dynamic loss scaling (init $2^{16}$,
floor applied; \S\ref{sec:divergences}). EMA decay 0.9999 with warmup ramp,
applied every step, including BN running statistics. \\
\midrule
\textbf{Augmentation} & Mosaic 1.0 (off for final 10 epochs);
MixUp 0.1~\citep{zhang2018mixup}; copy-paste 0.1~\citep{ghiasi2021copypaste};
HSV jitter (0.015, 0.7, 0.4); affine scale 0.9, translate 0.1; horizontal flip
0.5; random erasing 0.4~\citep{zhong2020randomerasing}; letterbox to
$640^2$. \\
\midrule
\textbf{Parallelism} & Data-parallel, one process per GPU, world 3; per-GPU
batch 16 (effective 48); bitwise-deterministic sharded all-reduce;
rank-invariant epoch shuffling with per-rank augmentation streams
(\S\ref{sec:divergences}); seed 0. \\
\midrule
\textbf{Hardware} & $3\times$ NVIDIA RTX 3090 (24\,GB), power-capped
200\,W each (graphics clock $\leq$1500\,MHz), consumer host (6 physical
cores); executed as $\approx$8-hour segments with automatic checkpoint-resume
($\approx$25 cycles) as a host-reliability mitigation. \\
\bottomrule
\end{tabularx}
\end{table}

\subsection{The reproduction protocol}\label{sec:method:protocol}

The reference implementation --- Ultralytics 8.3.75 on PyTorch
2.12.1~\citep{jocher2023yolov8,paszke2019pytorch} --- is treated as an
\emph{executable specification}: its configuration surface, numerical
conventions, and observable behavior define the target. This is the
differential-oracle arrangement of the companion
study~\citep{tran2026crossstack} with the roles fixed: the incumbent plays the
oracle, because the claim under test is that the independent stack reproduces
it. No reference source code is incorporated into numbat; parity is achieved
by independent implementation, validated at five levels. When any level
disagrees, numbat is changed to match the reference, never the reverse.

\textbf{L1 --- Operator.} Every differentiable operator passes a numeric
gradient check on CPU (tolerance $10^{-4}$) and a CPU-vs-CUDA forward/backward
equivalence check. Initializer distributions and the Philox RNG are verified
by known-answer tests against the reference (GPU uniform sampling is
bit-identical).

\textbf{L2 --- Module.} Composite components --- the detection loss with
task-aligned assignment, and batch normalization under autocast --- pass
known-answer tests against reference outputs on fixed inputs, on both CPU and
CUDA.

\textbf{L3 --- Step.} Fixed-data probes compare optimization \emph{kinetics}:
single- and multi-step loss decrease on identical batches, gradient
equivalence between 1-rank and $N$-rank execution (bitwise, by construction of
the deterministic all-reduce), and AMP-vs-F32 step-for-step agreement (4.2\%
mean absolute loss deviation over 285 steps after the fixes of
\S\ref{sec:divergences}). A \emph{kinetics decomposition} attributes any
trajectory difference to per-step progress vs.\ step count, which localized
several divergences below.

\textbf{L4 --- Trajectory.}\label{sec:method:gate} Before committing multi-day
compute, the reference implementation itself is run from scratch \emph{on the
same machine} at matched recipe and regime (3-GPU DDP, effective batch 48,
official augmentation) for 30 epochs, producing a reference trajectory
(Fig.~\ref{fig:convergence}a). The production run is then supervised by an
automated \emph{trajectory gate}: a sidecar polls the run every 10 minutes and
pauses or terminates it if the validation metric leaves a 20\% tolerance band
under the reference curve, plateaus, or diverges. Two operational rules
emerged and are now part of the protocol: (i) \emph{a reference band is valid
only for the exact recipe and regime it was recorded under} --- an earlier
band recorded with default (weaker) augmentation produced a spurious
``growing lag'' verdict against the strong-augmentation run; and (ii)
\emph{short probes of the reference must pin its optimizer explicitly}, since
its automatic optimizer selection silently overrides the configured learning
rate and momentum.

\textbf{L5 --- Outcome.}\label{sec:method:outcome} Reported accuracy comes
from the official COCO protocol only: numbat weights are exported to safetensors, loaded by the
\emph{reference stack's own validator}, and scored end-to-end with pycocotools
--- eliminating any possibility that numbat's metric implementation flatters
the result. numbat's internal streaming evaluator, used solely for gating, was
audited against this pipeline and reads 0.011--0.014 \emph{lower} than the
official score at converged checkpoints.

\section{Divergence catalog}\label{sec:divergences}

An early, un-gated attempt motivated the protocol: it silently plateaued at
roughly one fifth of the target metric for 43 epochs ($\approx$1.5
machine-days) before a human noticed --- no crash, no NaN, loss decreasing.
Post-mortem attributed it to a combination of the divergences below. All were
subsequently caught \emph{by the protocol}, fixed by adopting the reference
behavior, and re-validated; Table~\ref{tab:divergences} catalogs them.

\begin{table}[t]
\caption{Ten silent recipe divergences surfaced by the reproduction protocol.
Every fix adopted the reference behavior. ``Level'' = protocol level that
detected it (\S\ref{sec:method:protocol}).}
\label{tab:divergences}
\centering\footnotesize
\begin{tabularx}{\textwidth}{@{}c l X c@{}}
\toprule
\# & Component & Divergence $\rightarrow$ consequence & Level \\
\midrule
1 & Weight decay & Value taken from a released checkpoint's arguments was
already batch-scaled (batch 128); at effective batch 48 this doubled
regularization. Fix: re-derive from the base default via
$\lambda\, b w/\mathrm{nbs}$. & L4 \\
2 & AMP autocast & BatchNorm executed in the F16 shadow instead of F32
$\rightarrow$ degraded convergence and elevated overflow-skip rate. Fix: F32
BN and loss, F16 conv/matmul (reference autocast placement). & L2/L4 \\
3 & Initialization & Convolutions used Kaiming-normal (gain $\sqrt2$) instead
of the reference's Kaiming-uniform ($a{=}\sqrt5$)~\citep{he2015delving}, a
$2.45\times$ wider distribution; in the normalization-free detection head this inflated initial
classification-logit spread $\approx$14$\times$ and the resulting corrective
gradients overflowed F16. & L3 \\
4 & Loss scaler under DDP & With divergence \#2/\#3 present, per-rank overflow
flags OR-reduced across 3 ranks collapsed the dynamic loss scale toward zero
(silent learning freeze). Fix: scale floor; standard dynamics restored once
\#2/\#3 were fixed. & L4 \\
5 & Mosaic labels & Labels were not clipped to the mosaic canvas, producing
phantom boxes in padding regions. & L3 \\
6 & Optimizer groups & Normalization gains were placed in the elevated-warmup
bias group; the reference warms them from zero in their own group. & L4 \\
7 & Gradient clipping & Absent, vs.\ the reference's global-norm
10.0~\citep{pascanu2013difficulty} every step (a coded, not configured,
behavior). & L4 \\
8 & EMA cadence & EMA skipped on loss-scaler-rejected steps; the reference
updates unconditionally. & L4 \\
9 & Evaluation BN & Validation ran train-mode BN (batch statistics) on an EMA
model whose BN buffers were never averaged $\rightarrow$ noisy, low-biased
metric that both hides progress and fakes failure. Fix: eval-mode running
statistics, EMA including buffers. & L5 \\
10 & Augmentation RNG & All ranks drew identical augmentation streams
(effective augmentation diversity $\div$3). Fix: rank-folded batch seeds with
rank-invariant shuffling. A matched 5-epoch probe measured the single fix at
$+0.051$ mAP over the window vs.\ $+0.005$ without it. & L3/L4 \\
\bottomrule
\end{tabularx}
\end{table}

Three observations generalize beyond this case study. First,
\emph{divergences compose}: \#3 (initialization) amplified \#2 (autocast
placement), which triggered \#4 (scaler collapse) --- three individually
plausible implementations combining into a silent training freeze, while the
F32 control self-healed within seven steps and masked the chain. Second,
\emph{distinct root causes share one symptom}. Four different divergences
presented identically as ``slightly below the reference band from epoch 1,''
which is why a leveled protocol that can localize (operator? step kinetics?
recipe? metric?) terminates debugging that a single end-metric cannot. Third,
\emph{effective batch dominates short-horizon kinetics}: at effective batch 16,
\emph{both} frameworks make near-zero early progress under this recipe; the
reference's gradient accumulation to a nominal batch of 64 is part of the
recipe, not a tuning nicety --- probes that ignore it mislead.

\section{Results}\label{sec:results}

The run is complete: all numbers below are final, measured on the released
epoch-500 weights or recorded during the completed 500-epoch schedule.

\begin{figure}[t]
  \centering
  \includegraphics[width=\textwidth]{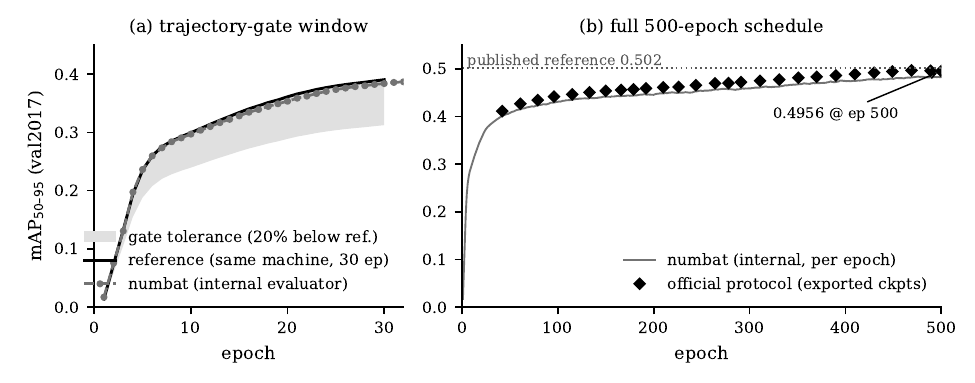}
  \caption{Convergence of the from-scratch run. (a) The trajectory-gate
  window: numbat's per-epoch validation metric tracks the same-machine
  reference trajectory (recorded with the reference implementation at matched
  recipe and regime) along the entire 30-epoch window; the automated gate
  polled 1{,}403 times over the completed run with zero trajectory
  violations. (b) The
  full schedule: internal per-epoch metric (line; reads 0.011--0.014 low,
  \S\ref{sec:method:protocol}~L5) and official-protocol scores of exported
  checkpoints (diamonds), against the published endpoint for the architecture
  class.}
  \label{fig:convergence}
\end{figure}

\subsection{Accuracy}\label{sec:results:acc}

Table~\ref{tab:accuracy} and Figure~\ref{fig:convergence} summarize accuracy.
The final (epoch-500) EMA checkpoint --- the released weights --- scores
\textbf{\FinalMap} mAP$_{50\text{--}95}$ (\FinalMapFifty{}
mAP$_{50}$) on COCO val2017 under the official protocol, scored by the
reference stack's validator; the published endpoint for the architecture class
at this scale and schedule is 0.502, placing the final score within 1.3\%
(relative) of it. The best official score observed during training was 0.4970
at epoch 469 (recorded in the released score trend). Notably, the final ten
mosaic-free epochs (\S\ref{sec:method}), which typically contribute a final
fraction of a point in this recipe family, produced no further gain in this
run: the official score moved 0.4970 (469) $\rightarrow$ 0.4962 (489)
$\rightarrow$ 0.4956 (500), differences at the scale of single-evaluation
noise. We report the epoch-500 number as the headline because it corresponds
to the released checkpoint; the epoch-469 value is reproducible from the
released trend.
Early-trajectory equivalence is direct: over the 30-epoch gate window numbat's
curve lies on the same-machine reference trajectory
(Fig.~\ref{fig:convergence}a), no lower than 97.9\% of the reference value at
any comparison epoch and up to $1.24\times$ above it in the earliest epochs,
\emph{before} correcting for the internal evaluator's known $-0.011$ to
$-0.014$ offset --- i.e., at or above the reference after correction.

\begin{table}[t]
\caption{COCO val2017 accuracy (official protocol; exported weights scored by
the reference validator + pycocotools). The published endpoint is the
reference implementation's reported from-scratch result for the architecture
class at this scale.}
\label{tab:accuracy}
\centering\small
\begin{tabular}{@{}lccc@{}}
\toprule
 & mAP$_{50\text{--}95}$ & mAP$_{50}$ & epochs \\
\midrule
\yolonbm{} final, released weights (this work) & \FinalMap & \FinalMapFifty &
500/500\\
\yolonbm{} best during training (this work) & 0.4970 & 0.6620 & 469 \\
Published endpoint (architecture class) & 0.502 & --- & 500 \\
Same-machine reference trajectory (gate baseline) & 0.391 & --- & 30 \\
\bottomrule
\end{tabular}
\end{table}

\begin{figure}[t]
  \centering
  \includegraphics[width=0.52\textwidth]{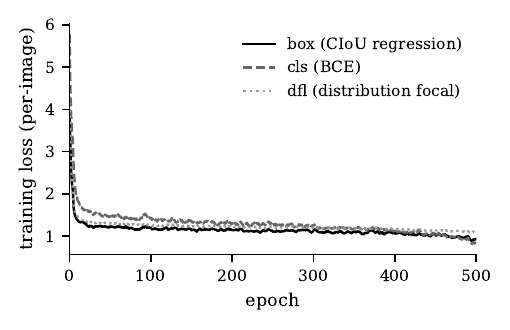}
  \caption{Training-loss components over the schedule (per-image, smoothed):
  CIoU box regression (\emph{box}), binary-cross-entropy classification
  (\emph{cls}), and distribution focal loss (\emph{dfl}).}
  \label{fig:losses}
\end{figure}

\subsection{Throughput, memory, energy}\label{sec:results:throughput}

\begin{figure}[t]
  \centering
  \includegraphics[width=\textwidth]{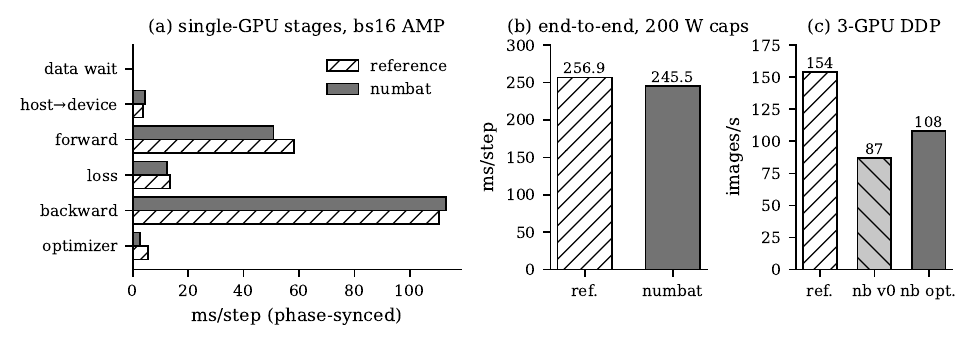}
  \caption{Same-machine training throughput vs.\ the reference at the
  production configuration (batch 16/GPU, AMP, $640^2$). (a) Phase-synchronized
  single-GPU stage times. (b) End-to-end step time under the run's permanent
  200\,W power caps; the numbat measurement includes live metric streaming,
  the reference runs with auxiliary outputs disabled. (c) Three-GPU
  data-parallel throughput: reference, numbat before and after the
  communication optimizations described in the text.}
  \label{fig:throughput}
\end{figure}

\emph{Single GPU: parity.} At the production configuration on one RTX 3090,
phase-synchronized stage timing (Fig.~\ref{fig:throughput}a) shows numbat
ahead on forward (50.8 vs.\ 58.2\,ms) and optimizer step (2.6 vs.\ 5.5\,ms),
at parity on backward (113.0 vs.\ 110.6\,ms) and loss (12.3 vs.\ 13.4\,ms,
after replacing a scalar-atomic reduction with a block reduction:
48$\rightarrow$12\,ms), with data loading fully overlapped by both systems.
End-to-end asynchronous step time is 185.9\,ms (numbat) vs.\ 192.8\,ms
(reference) uncapped, and 245.5 vs.\ 256.9\,ms under the run's permanent
200\,W caps --- with numbat additionally carrying live metric streaming
(Fig.~\ref{fig:throughput}b). We consider single-GPU training performance at
parity, established end-to-end at the production configuration rather than on
a favorable sub-benchmark.

\emph{Three-GPU DDP: honest gap.} At world size 3 the reference sustains
15.6\,min/epoch (154\,img/s) vs.\ numbat's initial 27.8\,min/epoch
(87\,img/s). Profiling attributes the gap to the serialized host-staged
all-reduce, which also absorbs the full inter-rank arrival skew inside the
step, whereas the reference overlaps bucketed NCCL reductions with
backpropagation~\citep{li2020pytorchddp}. Five optimizations --- pinned
staging (host-device copies $\approx$halved), sharding the reduction across
ranks (60--106$\rightarrow$19--26\,ms), futex barriers replacing spin-waits,
the fused gradient-norm kernel (\S\ref{sec:framework}), and folding the
loss-scaler overflow flag into the gradient payload (one fewer barrier round)
--- raised throughput to $\approx$108\,img/s with bitwise-identical gradient
math (Fig.~\ref{fig:throughput}c). Both remaining levers are implemented but
\emph{disabled by default on this host class}, a negative result we report
deliberately: on the 6-physical-core host, both the backward-overlapped
reduction and the NCCL transport starve the augmentation workers (data wait
0$\rightarrow$60--155\,ms/step), a net loss; they await validation on
larger-core hosts. Convergence is unaffected throughout: gradient reduction is
bitwise deterministic in all configurations.

\emph{Memory.} Peak training memory is 13.5\,GiB/GPU after incremental
activation freeing ($-26.8\%$ from 18.5), vs.\ $\approx$7\,GiB for the
reference at the same configuration; the residual factor $\approx$1.9$\times$
is attributed to AMP shadow parameter copies and allocator high-water
behavior, and is an open engineering item.

\emph{Energy and reliability.} Under the 200\,W caps, aggregate GPU board
power is bounded by 0.6\,kW; at the sustained 25--28\,min/epoch the 500-epoch
schedule completed in 9.6 days of wall clock including cool-down pauses,
bounding GPU energy at $\leq$1.4$\times$10$^{2}$\,kWh. The run executed as
$\approx$30 automatic checkpoint-resume segments with full state continuity
(parameters, optimizer, EMA, RNG); the trajectory gate
(Fig.~\ref{fig:convergence}a) polled 1{,}403 times over the run with zero
violations and zero human interventions after launch.

\section{Reproducibility and release}\label{sec:repro}

\begin{table}[t]
\caption{Environment of record.}
\label{tab:env}
\centering\small
\begin{tabular}{@{}ll@{}}
\toprule
GPUs & $3\times$ NVIDIA GeForce RTX 3090, 24\,GB, 200\,W cap, $\leq$1500\,MHz \\
Host & Intel Core i5-10600K (6C/12T), Debian 13 \\
numbat toolchain & Zig 0.17.0-dev (pinned); CUDA 12.9; driver 610.43.02 \\
Reference stack & Ultralytics 8.3.75; PyTorch 2.12.1+cu126; pycocotools \\
Dataset & COCO 2017 (train 118{,}287 / val 5{,}000)~\citep{lin2014coco} \\
Seed & 0 (data order, augmentation streams, initialization) \\
\bottomrule
\end{tabular}
\end{table}

We release: the trained weights (a safetensors export of the final EMA
checkpoint --- the exact artifact scored in \S\ref{sec:results}),
per-epoch training and validation metrics, the official-protocol score trend
of exported checkpoints, training-curve exports, and a manifest recording
every hyperparameter of Table~\ref{tab:recipe} with the environment of
Table~\ref{tab:env}, under a research-use-only (non-commercial) license.%
\footnote{\url{https://huggingface.co/cloudkites/yolo-nb-m}. Commercial
licensing: \texttt{contact@cloudkites.com}.}
The claim structure of this paper is independently verifiable without access
to numbat: the released safetensors load in the reference stack, and the
reported numbers regenerate from its validator and pycocotools.

The released weights are an original work: training started from random
initialization; no pretrained weights of any provenance were used; and the
training system shares no source code with any third-party machine-learning
framework. The reference implementation served only as an executable
behavioral specification and benchmark during development
(\S\ref{sec:method:protocol}) and as an independent scorer of the exported
weights. Verification of every reported number goes through the released
artifacts and the reference stack alone, and requires no access to numbat in
any form.

\section{Limitations}\label{sec:limits}

\emph{Single seed.} The headline run is one seed (0); a variance study at 500
epochs $\times$ 3 GPUs was outside our compute budget. The same-machine
reference baseline is likewise single-seed, so the equivalence claim is
trajectory- and endpoint-level, not a distributional statement.
\emph{Reference trajectory length.} The same-machine reference run covers 30
epochs; beyond it, equivalence rests on the official published endpoint and
the official-protocol scoring of our exported checkpoints.
\emph{Multi-GPU scaling.} The DDP gap of
\S\ref{sec:results:throughput} is real on small-core hosts; the implemented
overlap and NCCL paths remain to be validated on larger hosts.
\emph{Memory.} numbat currently uses $\approx$1.9$\times$ the reference's
training memory at this configuration.
\emph{Scope.} The protocol is demonstrated on one detector family; the
framework's LLM/ASR/segmentation stacks run in production but have not yet
been taken through the full five-level protocol.
\emph{Availability.} The framework's source is proprietary, and the terms
under which the SDK is distributed are a commercial matter outside this
paper's scope. The paper is written so that neither is needed: every reported
score regenerates from the released weights, the reference stack's validator
and pycocotools, and the claim structure stands or falls on those public
artifacts.

\section{Conclusion}\label{sec:conclusion}

We presented numbat, a self-contained machine-learning stack, and the
engineering discipline that holds it to account: reference-default semantics so
recipes transfer unchanged, an additive C ABI so applications outlive
framework releases, domain requirements encoded as executable acceptance
gates, and a five-level reproduction protocol that treats the incumbent
implementation as an executable specification. The validation is the
strongest test we know for a training stack: reproducing a mature,
competitive recipe's outcome from scratch, with the evidence scored by the
incumbent's own tooling --- and, in the companion study, the same two-stack
arrangement run in the opposite direction, as a fault-finding
instrument~\citep{tran2026crossstack}. Between them the exercises yield two
transferable artifacts, the divergence catalog and the protocol itself, and
one existence proof: training-outcome parity with the dominant ecosystem is
achievable in a fully independent stack, on consumer hardware. Future work
extends the protocol to the framework's language-model and speech stacks,
closes the multi-GPU overlap and memory gaps, and carries the acceptance-gate
method into the domain planes still under construction, where we expect it to
keep earning its keep the way it has so far: by refusing to let a passing
build stand in for a discharged requirement.

\section*{List of abbreviations}

\footnotesize
\noindent\begin{tabular}{@{}ll@{\qquad}ll@{}}
ABI & application binary interface & FPN & feature-pyramid network \\
AMP & automatic mixed precision & IoU & intersection over union \\
BCE & binary cross-entropy & JIT & just-in-time (compilation) \\
BF16 & bfloat16 floating point & LLM & large language model \\
BLAS & Basic Linear Algebra Subprograms & mAP & mean average precision \\
BN & batch normalization & nbs & nominal batch size \\
CIoU & complete intersection over union & NCCL & NVIDIA Collective Communications Library \\
CSP & cross-stage partial (backbone) & NVRTC & NVIDIA runtime compiler \\
CTC & connectionist temporal classification & OCR & optical character recognition \\
DAG & directed acyclic graph & PAN & path-aggregation network \\
DDP & distributed data parallelism & RNG & random-number generator \\
DFL & distribution focal loss & SDK & software development kit \\
EMA & exponential moving average & SDPA & scaled dot-product attention \\
F16/F32/F64 & 16/32/64-bit floating point & SGD & stochastic gradient descent \\
 & & SIMD & single instruction, multiple data \\
 & & VLM & vision--language model \\
 & & WGSL & WebGPU shading language \\
\multicolumn{4}{@{}l}{DICOM \; Digital Imaging and Communications in Medicine} \\
\end{tabular}
\normalsize

\bibliographystyle{plainnat}
\bibliography{references}

@inproceedings{paszke2019pytorch,
  title     = {{PyTorch}: An Imperative Style, High-Performance Deep Learning Library},
  author    = {Paszke, Adam and Gross, Sam and Massa, Francisco and Lerer, Adam and Bradbury, James and Chanan, Gregory and Killeen, Trevor and Lin, Zeming and Gimelshein, Natalia and Antiga, Luca and Desmaison, Alban and K{\"o}pf, Andreas and Yang, Edward and DeVito, Zachary and Raison, Martin and Tejani, Alykhan and Chilamkurthy, Sasank and Steiner, Benoit and Fang, Lu and Bai, Junjie and Chintala, Soumith},
  booktitle = {Advances in Neural Information Processing Systems 32},
  year      = {2019}
}

@inproceedings{ansel2024pytorch2,
  title     = {{PyTorch}~2: Faster Machine Learning Through Dynamic {Python} Bytecode Transformation and Graph Compilation},
  author    = {Ansel, Jason and Yang, Edward and He, Horace and Gimelshein, Natalia and Jain, Animesh and Voznesensky, Michael and others},
  booktitle = {29th ACM International Conference on Architectural Support for Programming Languages and Operating Systems (ASPLOS)},
  pages     = {929--947},
  year      = {2024}
}

@inproceedings{abadi2016tensorflow,
  title     = {{TensorFlow}: A System for Large-Scale Machine Learning},
  author    = {Abadi, Mart{\'i}n and Barham, Paul and Chen, Jianmin and Chen, Zhifeng and Davis, Andy and Dean, Jeffrey and Devin, Matthieu and Ghemawat, Sanjay and Irving, Geoffrey and Isard, Michael and others},
  booktitle = {12th USENIX Symposium on Operating Systems Design and Implementation (OSDI)},
  year      = {2016}
}

@misc{bradbury2018jax,
  title  = {{JAX}: Composable Transformations of {Python}+{NumPy} Programs},
  author = {Bradbury, James and Frostig, Roy and Hawkins, Peter and Johnson, Matthew James and Katariya, Yash and Leary, Chris and Maclaurin, Dougal and Necula, George and Paszke, Adam and VanderPlas, Jake and Wanderman-Milne, Skye and Zhang, Qiao},
  year   = {2018},
  note   = {\url{https://github.com/jax-ml/jax}}
}

@misc{gerganov2023llamacpp,
  title  = {llama.cpp: {LLM} Inference in {C}/{C}++},
  author = {Gerganov, Georgi and {contributors}},
  year   = {2023},
  note   = {\url{https://github.com/ggml-org/llama.cpp}}
}

@misc{onnxruntime2018,
  title  = {{ONNX Runtime}},
  author = {{ONNX Runtime developers}},
  year   = {2018},
  note   = {\url{https://onnxruntime.ai}}
}

@misc{burn2024,
  title  = {Burn: A Next-Generation Tensor Library and Deep Learning Framework},
  author = {Simard, Nathaniel and Fortier-Dubois, Louis and Tadjibaev, Dilshod and Lagrange, Guillaume and {contributors}},
  year   = {2024},
  note   = {\url{https://github.com/tracel-ai/burn}}
}

@misc{tinygrad2023,
  title  = {tinygrad},
  author = {Hotz, George and {the tiny corp}},
  year   = {2023},
  note   = {\url{https://github.com/tinygrad/tinygrad}}
}

@misc{hannun2023mlx,
  title  = {{MLX}: Efficient and Flexible Machine Learning on {Apple} Silicon},
  author = {Hannun, Awni and Digani, Jagrit and Katharopoulos, Angelos and Collobert, Ronan},
  year   = {2023},
  note   = {\url{https://github.com/ml-explore/mlx}}
}

@inproceedings{redmon2016yolo,
  title     = {You Only Look Once: Unified, Real-Time Object Detection},
  author    = {Redmon, Joseph and Divvala, Santosh and Girshick, Ross and Farhadi, Ali},
  booktitle = {IEEE Conference on Computer Vision and Pattern Recognition (CVPR)},
  year      = {2016}
}

@article{redmon2018yolov3,
  title   = {{YOLOv3}: An Incremental Improvement},
  author  = {Redmon, Joseph and Farhadi, Ali},
  journal = {arXiv preprint arXiv:1804.02767},
  year    = {2018}
}

@article{bochkovskiy2020yolov4,
  title   = {{YOLOv4}: Optimal Speed and Accuracy of Object Detection},
  author  = {Bochkovskiy, Alexey and Wang, Chien-Yao and Liao, Hong-Yuan Mark},
  journal = {arXiv preprint arXiv:2004.10934},
  year    = {2020}
}

@article{li2022yolov6,
  title   = {{YOLOv6}: A Single-Stage Object Detection Framework for Industrial Applications},
  author  = {Li, Chuyi and Li, Lulu and Jiang, Hongliang and Weng, Kaiheng and Geng, Yifei and Li, Liang and Ke, Zaidan and Li, Qingyuan and Cheng, Meng and Nie, Weiqiang and others},
  journal = {arXiv preprint arXiv:2209.02976},
  year    = {2022}
}

@inproceedings{wang2023yolov7,
  title     = {{YOLOv7}: Trainable Bag-of-Freebies Sets New State-of-the-Art for Real-Time Object Detectors},
  author    = {Wang, Chien-Yao and Bochkovskiy, Alexey and Liao, Hong-Yuan Mark},
  booktitle = {IEEE/CVF Conference on Computer Vision and Pattern Recognition (CVPR)},
  year      = {2023}
}

@misc{jocher2023yolov8,
  title  = {Ultralytics {YOLOv8}},
  author = {Jocher, Glenn and Chaurasia, Ayush and Qiu, Jing},
  year   = {2023},
  note   = {Software, version 8.x. \url{https://github.com/ultralytics/ultralytics}}
}

@inproceedings{wang2020cspnet,
  title     = {{CSPNet}: A New Backbone That Can Enhance Learning Capability of {CNN}},
  author    = {Wang, Chien-Yao and Liao, Hong-Yuan Mark and Wu, Yueh-Hua and Chen, Ping-Yang and Hsieh, Jun-Wei and Yeh, I-Hau},
  booktitle = {IEEE/CVF Conference on Computer Vision and Pattern Recognition Workshops},
  year      = {2020}
}

@inproceedings{lin2017fpn,
  title     = {Feature Pyramid Networks for Object Detection},
  author    = {Lin, Tsung-Yi and Doll{\'a}r, Piotr and Girshick, Ross and He, Kaiming and Hariharan, Bharath and Belongie, Serge},
  booktitle = {IEEE Conference on Computer Vision and Pattern Recognition (CVPR)},
  year      = {2017}
}

@inproceedings{liu2018panet,
  title     = {Path Aggregation Network for Instance Segmentation},
  author    = {Liu, Shu and Qi, Lu and Qin, Haifang and Shi, Jianping and Jia, Jiaya},
  booktitle = {IEEE Conference on Computer Vision and Pattern Recognition (CVPR)},
  year      = {2018}
}

@inproceedings{li2020gfl,
  title     = {Generalized Focal Loss: Learning Qualified and Distributed Bounding Boxes for Dense Object Detection},
  author    = {Li, Xiang and Wang, Wenhai and Wu, Lijun and Chen, Shuo and Hu, Xiaolin and Li, Jun and Tang, Jinhui and Yang, Jian},
  booktitle = {Advances in Neural Information Processing Systems 33},
  year      = {2020}
}

@inproceedings{feng2021tood,
  title     = {{TOOD}: Task-Aligned One-Stage Object Detection},
  author    = {Feng, Chengjian and Zhong, Yujie and Gao, Yu and Scott, Matthew R. and Huang, Weilin},
  booktitle = {IEEE/CVF International Conference on Computer Vision (ICCV)},
  year      = {2021}
}

@inproceedings{zheng2020diou,
  title     = {Distance-{IoU} Loss: Faster and Better Learning for Bounding Box Regression},
  author    = {Zheng, Zhaohui and Wang, Ping and Liu, Wei and Li, Jinze and Ye, Rongguang and Ren, Dongwei},
  booktitle = {AAAI Conference on Artificial Intelligence},
  year      = {2020}
}

@inproceedings{shen2017dsod,
  title     = {{DSOD}: Learning Deeply Supervised Object Detectors from Scratch},
  author    = {Shen, Zhiqiang and Liu, Zhuang and Li, Jianguo and Jiang, Yu-Gang and Chen, Yurong and Xue, Xiangyang},
  booktitle = {IEEE International Conference on Computer Vision (ICCV)},
  year      = {2017}
}

@inproceedings{he2019rethinking,
  title     = {Rethinking {ImageNet} Pre-Training},
  author    = {He, Kaiming and Girshick, Ross and Doll{\'a}r, Piotr},
  booktitle = {IEEE/CVF International Conference on Computer Vision (ICCV)},
  year      = {2019}
}

@article{pineau2021reproducibility,
  title   = {Improving Reproducibility in Machine Learning Research (A Report from the {NeurIPS} 2019 Reproducibility Program)},
  author  = {Pineau, Joelle and Vincent-Lamarre, Philippe and Sinha, Koustuv and Larivi{\`e}re, Vincent and Beygelzimer, Alina and d'Alch{\'e}-Buc, Florence and Fox, Emily and Larochelle, Hugo},
  journal = {Journal of Machine Learning Research},
  volume  = {22},
  number  = {164},
  pages   = {1--20},
  year    = {2021}
}

@inproceedings{henderson2018deeprl,
  title     = {Deep Reinforcement Learning That Matters},
  author    = {Henderson, Peter and Islam, Riashat and Bachman, Philip and Pineau, Joelle and Precup, Doina and Meger, David},
  booktitle = {AAAI Conference on Artificial Intelligence},
  year      = {2018}
}

@inproceedings{bouthillier2019unreproducible,
  title     = {Unreproducible Research is Reproducible},
  author    = {Bouthillier, Xavier and Laurent, C{\'e}sar and Vincent, Pascal},
  booktitle = {International Conference on Machine Learning (ICML), PMLR 97},
  pages     = {725--734},
  year      = {2019}
}

@inproceedings{salmon2011philox,
  title     = {Parallel Random Numbers: As Easy as 1, 2, 3},
  author    = {Salmon, John K. and Moraes, Mark A. and Dror, Ron O. and Shaw, David E.},
  booktitle = {International Conference for High Performance Computing, Networking, Storage and Analysis (SC)},
  year      = {2011}
}

@inproceedings{lavin2016winograd,
  title     = {Fast Algorithms for Convolutional Neural Networks},
  author    = {Lavin, Andrew and Gray, Scott},
  booktitle = {IEEE Conference on Computer Vision and Pattern Recognition (CVPR)},
  year      = {2016}
}

@inproceedings{tillet2019triton,
  title     = {Triton: An Intermediate Language and Compiler for Tiled Neural Network Computations},
  author    = {Tillet, Philippe and Kung, H. T. and Cox, David},
  booktitle = {3rd ACM SIGPLAN International Workshop on Machine Learning and Programming Languages (MAPL)},
  year      = {2019}
}

@inproceedings{micikevicius2018mixed,
  title     = {Mixed Precision Training},
  author    = {Micikevicius, Paulius and Narang, Sharan and Alben, Jonah and Diamos, Gregory and Elsen, Erich and Garcia, David and Ginsburg, Boris and Houston, Michael and Kuchaiev, Oleksii and Venkatesh, Ganesh and Wu, Hao},
  booktitle = {International Conference on Learning Representations (ICLR)},
  year      = {2018}
}

@article{li2020pytorchddp,
  title   = {{PyTorch} Distributed: Experiences on Accelerating Data Parallel Training},
  author  = {Li, Shen and Zhao, Yanli and Varma, Rohan and Salpekar, Omkar and Noordhuis, Pieter and Li, Teng and Paszke, Adam and Smith, Jeff and Vaughan, Brian and Damania, Pritam and Chintala, Soumith},
  journal = {Proceedings of the VLDB Endowment},
  volume  = {13},
  number  = {12},
  pages   = {3005--3018},
  year    = {2020}
}

@inproceedings{lin2014coco,
  title     = {Microsoft {COCO}: Common Objects in Context},
  author    = {Lin, Tsung-Yi and Maire, Michael and Belongie, Serge and Hays, James and Perona, Pietro and Ramanan, Deva and Doll{\'a}r, Piotr and Zitnick, C. Lawrence},
  booktitle = {European Conference on Computer Vision (ECCV)},
  year      = {2014}
}

@article{elfwing2018silu,
  title   = {Sigmoid-Weighted Linear Units for Neural Network Function Approximation in Reinforcement Learning},
  author  = {Elfwing, Stefan and Uchibe, Eiji and Doya, Kenji},
  journal = {Neural Networks},
  volume  = {107},
  pages   = {3--11},
  year    = {2018}
}

@inproceedings{zhang2018mixup,
  title     = {mixup: Beyond Empirical Risk Minimization},
  author    = {Zhang, Hongyi and Ciss{\'e}, Moustapha and Dauphin, Yann N. and Lopez-Paz, David},
  booktitle = {International Conference on Learning Representations (ICLR)},
  year      = {2018}
}

@inproceedings{ghiasi2021copypaste,
  title     = {Simple Copy-Paste is a Strong Data Augmentation Method for Instance Segmentation},
  author    = {Ghiasi, Golnaz and Cui, Yin and Srinivas, Aravind and Qian, Rui and Lin, Tsung-Yi and Cubuk, Ekin D. and Le, Quoc V. and Zoph, Barret},
  booktitle = {IEEE/CVF Conference on Computer Vision and Pattern Recognition (CVPR)},
  year      = {2021}
}

@inproceedings{zhong2020randomerasing,
  title     = {Random Erasing Data Augmentation},
  author    = {Zhong, Zhun and Zheng, Liang and Kang, Guoliang and Li, Shaozi and Yang, Yi},
  booktitle = {AAAI Conference on Artificial Intelligence},
  year      = {2020}
}

@inproceedings{he2015delving,
  title     = {Delving Deep into Rectifiers: Surpassing Human-Level Performance on {ImageNet} Classification},
  author    = {He, Kaiming and Zhang, Xiangyu and Ren, Shaoqing and Sun, Jian},
  booktitle = {IEEE International Conference on Computer Vision (ICCV)},
  year      = {2015}
}

@inproceedings{ioffe2015batch,
  title     = {Batch Normalization: Accelerating Deep Network Training by Reducing Internal Covariate Shift},
  author    = {Ioffe, Sergey and Szegedy, Christian},
  booktitle = {International Conference on Machine Learning (ICML)},
  year      = {2015}
}

@inproceedings{pascanu2013difficulty,
  title     = {On the Difficulty of Training Recurrent Neural Networks},
  author    = {Pascanu, Razvan and Mikolov, Tomas and Bengio, Yoshua},
  booktitle = {International Conference on Machine Learning (ICML)},
  year      = {2013}
}

@article{tran2026crossstack,
  title   = {Cross-Stack Validation of Language-Model Training: A Clinical Fine-Tuning Case Study},
  author  = {Tran, Thang and Dang, Lan},
  journal = {arXiv preprint arXiv:2608.24267},
  year    = {2026}
}

\end{document}